\documentclass[journal]{IEEEtran} 

\usepackage{cite} 
\usepackage{amsmath,amssymb,amsfonts} 
\usepackage{algorithm}  
\usepackage{algorithmic}
\usepackage{array}  
\usepackage[caption=false,font=normalsize,labelfont=rm,textfont=rm]{subfig}
\usepackage{textcomp}
\usepackage{stfloats}
\usepackage{url}  
\usepackage{bm}
\usepackage{makecell}
\usepackage{verbatim}
\usepackage{booktabs}
\usepackage{graphicx}  
\usepackage{multirow}  
\usepackage{caption}
\usepackage{hyperref} 
\def\BibTeX{{\rm B\kern-.05em{\sc i\kern-.025em b}\kern-.08em
    T\kern-.1667em\lower.7ex\hbox{E}\kern-.125emX}}
\usepackage{balance}

\begin{document}
\title{Adaptive Wireless Image Semantic Transmission: Design, Simulation, and Prototype Validation}

\author{Jiarun Ding, Peiwen Jiang, \textit{Graduate Student Member}, \textit{IEEE}, Chao-Kai Wen, \textit{Fellow}, \textit{IEEE} and\\ Shi Jin, \textit{Fellow}, \textit{IEEE}
\thanks{J. Ding, P. Jiang and S. Jin are with the National Mobile Communications Research Laboratory, Southeast University, Nanjing 210096, China (e-mail: jrunding@seu.edu.cn; PeiwenJiang@seu.edu.cn; jinshi@seu.edu.cn).\par
C.-K. Wen is with the Institute of Communications Engineering, National Sun Yat-sen University, Kaohsiung 80424, Taiwan (e-mail: chaokai.wen@mail.nsysu.edu.tw).
}}

\maketitle
\pagestyle{empty}  % no page number for the second and the later pages
\thispagestyle{empty} % no page number for the first page

\begin{abstract}
The rapid development of artificial intelligence has significantly advanced semantic communications, particularly in wireless image transmission. However, most existing approaches struggle to precisely distinguish and prioritize image content, and they do not sufficiently incorporate semantic priorities into system design. In this study, we propose an adaptive wireless image semantic transmission scheme called ASCViT-JSCC, which utilizes vision transformer-based joint source-channel coding (JSCC). This scheme prioritizes different image regions based on their importance, identified through object and feature point detection. Unimportant background sections are masked, enabling them to be recovered at the receiver, while the freed resources are allocated to enhance object protection via the JSCC network. We also integrate quantization modules to enable compatibility with quadrature amplitude modulation, commonly used in modern wireless communications. To address frequency-selective fading channels, we introduce CSIPA-Net, which allocates power based on channel information, further improving performance. Notably, we conduct over-the-air testing on a prototype platform composed of a software-defined radio and embedded graphics processing unit systems, validating our methods. Both simulations and real-world measurements demonstrate that ASCViT-JSCC effectively prioritizes object protection according to channel conditions, significantly enhancing image reconstruction quality, especially in challenging channel environments.
\end{abstract}


%%%%%%%%%%%%%%%% INTRODUCTION %%%%%%%%%%%%%%%%%%%%%%
\section{Introduction}
\IEEEPARstart{T}{he} upcoming sixth-generation (6G) communication framework is accelerating the integration of artificial intelligence (AI) into communication systems \cite{1}. By leveraging AI's powerful capabilities in semantic extraction, semantic communication---a paradigm initially introduced by Shannon and Weaver in 1949 \cite{3}---has gained increasing attention in recent research \cite{61}. Unlike traditional communication, which is based on classical information theory \cite{4} and focuses on accurate transmission of bits, semantic communication prioritizes  reliable and efficient transmission of semantic meanings behind bits \cite{5}. This shift towards semantics opens up new opportunities in applications such as extended reality \cite{7} and edge intelligence \cite{10}.

Currently, most cutting-edge research on semantic communication employs deep learning (DL)-based joint source-channel coding (JSCC) to implement encoders and decoders (codecs) for multimedia data transmission, such as text \cite{18,74}, audio \cite{73,21}, image \cite{60,16} and video \cite{23,15}. This paradigm contrasts with the separate source-channel coding (SSCC) approach based on Shannon’s classical information theory. While SSCC is optimal for memoryless sources and channels under ideal conditions of unconstrained latency, complexity, and code length \cite{4}, such conditions are rarely met in practical systems. As a result, SSCC is often suboptimal, making JSCC-based semantic communications more competitive in various scenarios.

Among multimedia data, image and video serve as critical carriers of information, rich in semantics. Consequently, they have been extensively studied within the context of JSCC-based semantic communications. Early work, such as DeepJSCC \cite{14}, which leveraged deep neural networks (NNs) for image transmission, directly mapped input images to channel symbols and enabled the decoder to reconstruct the images from distorted symbols. Further study has explored video conferencing \cite{24}, which reduces bandwidth usage by transmitting only keypoints during video conferencing. Various improvements to this architecture, such as adaptive rate control based on channel conditions \cite{34} and the use of OFDM for multipath fading channels \cite{36}, have been proposed. Additionally, the incorporation of channel state information (CSI) \cite{37} and attention mechanism \cite{39,40} have also enhanced the performance of JSCC-based image transmission. However, despite these advancements, existing systems struggle to differentiate and prioritize important image content, limiting the flexibility of communication systems in managing diverse semantic information and adapting to dynamic wireless channel conditions.

A significant challenge facing current semantic communication systems lies in their lack of generalization. Since many networks are trained under specific conditions, their trained parameters often fail to perform optimally when these conditions change. Foundation models (FMs), which are known for their robust semantic extraction and generation capabilities, present a potential solution to this challenge, offering new avenues for improving generalization in semantic communications. For instance, the FM-based video conference framework Txt2Vid \cite{69} showcased a substantial reduction in bandwidth through compressing videos to text transcripts, while the large AI model-based semantic communication framework (LAM-SC) \cite{64} integrated a large model with an explicit knowledge base to enhance generalization. Other studies have also explored the role of FMs across various system layers, taking into account computational complexity and task-specific applications \cite{71,72}. By integrating pre-trained models into communication systems, data compression can be maximized, thus minimizing the volume of transmitted data---a key insight that inspired the development of our proposed system.

Although JSCC-based semantic communications combined with FMs can address many issues related to semantic extraction, generation, and generalization, simply incorporating these methods does not fully leverage the advantages of semantic communications to enhance performance in physical layer transmission. As a result, physical layer challenges remain, particularly in adapting to dynamic wireless channel environments \cite{6}. Several studies have explored integrating semantic communications with physical layer designs, such as constellation design \cite{27} and peak-to-average power ratio reduction \cite{28}. Nan \textit{et al.} also studied physical-layer adversarial robustness of semantic communications to enhance the physical-layer security \cite{70}. However, the impact of semantics on overcoming frequency-selective fading has not been adequately addressed, which is one of the problems this work seeks to tackle.

Additionally, many of existing works in semantic communications remain at simulation stage, and their results often lack practical relevance for real-world applications. Bridging this gap requires prototype validation, which is crucial for transitioning from simulations to actual applications, especially in the context of intelligent communications such as semantic communications. Although some efforts have been made in this direction, further progress is needed. For example, the 5G-compliant RaPro system \cite{45} combined FPGA-privileged modules from software-defined radio (SDR) with high-level programming languages on multi-core general-purpose processors (GPPs). In \cite{46}, an AI-aided OFDM receiver was implemented using C/C++ on RaPro, but deploying intelligent algorithms proved difficult due to the inherent complexity of C/C++. Another testbed proposed in \cite{48} used USRP-2943R and host PCs to develop a wireless semantic communication system, but it was based on analog and single-carrier communication, differing from modern wideband wireless communication systems. Therefore, there is a clear need for further development of prototype validation platforms for semantic communications.

To overcome the aforementioned challenges, we propose a novel adaptive wireless image semantic transmission scheme, ASCViT-JSCC, which adheres to digital modulation standards. Our scheme incorporates YOLOv5 \cite{49} and the scale-invariant feature transform (SIFT) \cite{50} to differentiate between objects and backgrounds in images, selectively masking background areas based on the current channel conditions. The JSCC NN then prioritizes the preservation of important objects with the assistance of these masked regions through end-to-end training. On the receiving side, the masked backgrounds are reconstructed using a masked autoencoder (MAE) decoder \cite{51}. Furthermore, we have designed and implemented a prototype validation platform, the Intelligent Communication Prototype (ICP), using USRP-2943R and NVIDIA Jetson Xavier NX to evaluate and compare our proposed scheme with existing methods, thereby verifying its feasibility and superiority.

The major contributions of this paper are summarized as follows:
\begin{enumerate} 
\item We propose an adaptive wireless image semantic transmission method that adjusts to both channel conditions and user requirements. The method uses YOLOv5 and SIFT to analyze priority in different parts of images. A vision transformer (ViT)-based \cite{52} JSCC NN, equipped with parameterless quantization modules, is designed for wireless digital image transmission, enhancing the protection of high-priority objects. 

\item In frequency-selective channels, we introduce a robust fully-connected network (FCN)-based scheme, CSIPA-Net, which reallocates power across subchannels based on OFDM channel conditions. CSIPA-Net assigns more power to subchannels with higher SNRs to preserve key image features. Experimental results show that ASCViT-JSCC with CSIPA-Net significantly outperforms implementations without CSIPA-Net. 

\item We develop a OFDM prototype validation platform named ICP. Practical measurements and performance analyses are presented, demonstrating the feasibility of the proposed scheme. We also evaluate the complexity of our method and outline potential future improvements. 
\end{enumerate} 

The remainder of this paper is organized as follows: Section \uppercase\expandafter{\romannumeral2} covers the system model and transceiver design of ASCViT-JSCC. Section \uppercase\expandafter{\romannumeral3} presents simulation results. Section \uppercase\expandafter{\romannumeral4}  details the design and evaluation of the ICP platform, and Section \uppercase\expandafter{\romannumeral5} concludes with suggestions for future improvements.

%%%%%%%%  2 TRANSCEIVER DESIGN %%%%%%%%%%%%
\section{System Model and Transceiver Design}
In this section, we first introduce the system model of the proposed ASCViT-JSCC framework. Next, we describe the key modules, including adaptive preprocessing and the JSCC NN architecture used in ASCViT-JSCC. Finally, we present the design of a robust and independent module, CSIPA-Net, specifically developed to counter the effects of frequency-selective fading channels.

\subsection{System Model}
\begin{figure*}[!t]
    \hspace{2mm}
	\centerline{\includegraphics[width=5.7in]{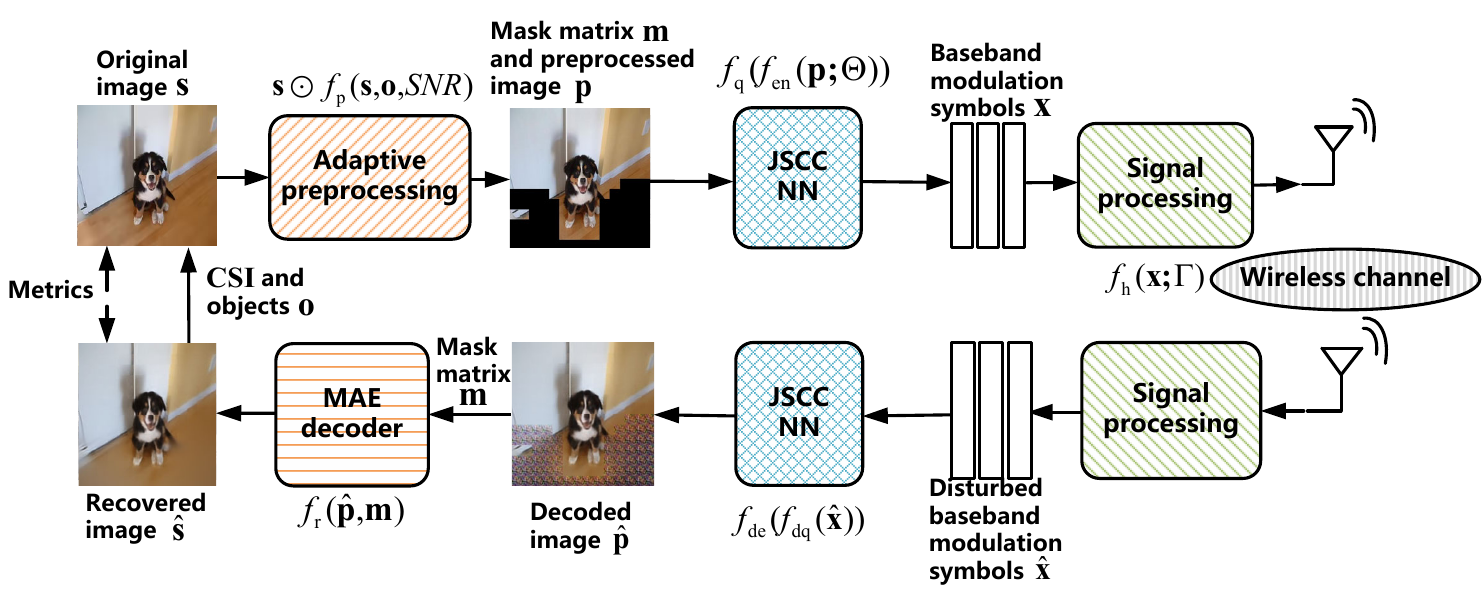}}
        \caption{The structure of ASCViT-JSCC. The upper part represents the transmitter, the lower part the receiver, and the right side depicts the wireless channel.}  
	\label{fig1}
    \vspace{-5mm}
\end{figure*}

The ASCViT-JSCC system, as shown in Fig. \ref{fig1}, consists of three main components: the transmitter, the wireless channel, and the receiver. The transmitter incorporates two key modules: the adaptive preprocessing module and the JSCC NN module. The adaptive preprocessing module generates a binary mask matrix based on the original image, channel conditions, and user requirements. This mask is used to prioritize important parts of the image. The ViT-based JSCC NN module then encodes the preprocessed image into baseband in-phase/quadrature (I/Q) data, referred to as semantic modulation data. After the transmission over the wireless channel, an asymmetric JSCC NN module at the receiver decodes the disturbed semantic modulation data into an impaired masked image. The final recovered image is synthesized by a MAE decoder. Additional modules such as channel estimation, signal detection, and the MAE decoder are common techniques integrated into the system.

Assume that the receiver performs channel estimation and feeds the CSI and the prioritized objects back to the transmitter. Let CSI and the desired objects be represented as $\textbf{CSI}\in{\mathbb{C}^{1\times{(K+1)}}}$ and $\textbf{o}\in{\mathbb{R}^{1\times{O}}}$, where $K$ is the number of subcarriers in OFDM, and $O$ is the number of prioritized objects. The vector $\textbf{CSI}$ contains subchannel gains and the overall signal-to-noise ratio (SNR). Objects are represented by integer indices (from 1 to 80) according to the COCO dataset \cite{54}. For example, if the receiver prioritizes ``dog'' objects, the transmitter detects these objects in the image and protects them accordingly.
 
Let the original image be denoted as $\textbf{s}\in{\mathbb{R}^{H\times{W}\times{C}}}$, where $H$, $W$ and $C$  represent the image's height, width, and the number of color channels, respectively. A mask matrix, $\textbf{m} \in{\mathbb{R}^{H\times{W}\times{C}}}$, is generated based on the prioritized objects $\textbf{o}$ and the SNR value, as described by the function:  
\begin{equation}
\textbf{m}=f_{\rm{p}}(\textbf{s}, \textbf{o}, \textit{SNR}),  \label{eq1} 
\end{equation}
where $f_{\rm{p}}(\cdot)$ is the function that generates the mask matrix. The original image $\textbf{s}$ is then masked by $\textbf{m}$, producing the preprocessed image $\textbf{p}\in{\mathbb{R}^{H\times{W}\times{C}}}$. In this preprocessed image, unimportant regions are masked, leaving only the desired objects and relevant background intact. 

Next, the JSCC NN encoder, denoted as $f_{\rm{en}}(\cdot \, ;\bm{\Theta})$, where $\bm{\Theta}$ represents the model parameters, encodes the preprocessed image into semantic floating-point data. This data is quantized by a non-trainable quantization module $f_{\rm{q}}(\cdot)$, converting it into semantic modulation data. The semantic modulation data is then paired and processed into baseband modulation symbols $\textbf{x}\in{\mathbb{R}^{1\times{\frac{N}{2}}}}$, where $\textit{N} $ is the total number of I/Q baseband data points. This process is described by 
%\vspace{-1mm}
\begin{equation}
\textbf{x}=f_{\rm{_q}}(f_{\rm{en}}(\textbf{s}\odot\textbf{m};\bm{\Theta})), 
\label{eq2} 
\end{equation}
where $\odot$ denotes element-wise multiplication.

The baseband modulation symbols are converted into OFDM waveforms and transmitted over the wireless channel. At the receiver, the received waveforms, distorted by the wireless channel, are processed into impaired baseband modulation symbols $\hat{\textbf{x}}\in{\mathbb{R}^{1\times{N}}}$ through channel estimation and signal detection. This is expressed as 
\begin{equation}
\hat{\textbf{x}} = f_{\rm{h}}(\textbf{x};\bm{\Gamma}),
\label{eq3} 
\end{equation}
where $f_{\rm{h}}(\cdot)$ represents the wireless channel effect, and $\bm{\Gamma}$ denotes the channel parameters.

At the receiver, a dequantization module $f_{\rm{dq}}(\cdot)$ embedded in the JSCC NN recovers the I/Q data from the impaired baseband modulation symbols. The JSCC NN decoder $f_{\rm{de}}(\cdot)$ then decodes the impaired semantic modulation data into a disturbed masked image $\hat{\textbf{p}}\in{\mathbb{R}^{H\times{W}\times{C}}}$. In this image, the desired objects are well protected, while the background, treated as redundant information, helps in preserving the overall image quality. Finally, a pre-trained MAE decoder $f_{\rm{r}}(\cdot \, ;\bm{\Psi})$ \cite{51} reconstructs the entire image $\hat{\textbf{s}}$, with $\bm{\Psi}$ representing the fixed parameters of the decoder. This process is represented by
\begin{equation} 
\hat{\textbf{s}}=f_{\rm{r}}(f_{\rm{de}}(f_{\rm{dq}}(\hat{\textbf{x}});\bm{\Phi}),\textbf{m};\bm{\Psi}), \label{eq4} 
\end{equation}
where $\bm{\Phi}$ represents the parameters of the JSCC NN at the receiver.

The overall objective of this system is to design an adaptive preprocessing module for efficient object extraction and preservation, and to determine the optimal JSCC NN parameters that minimize the reconstruction error between $\textbf{p}$ and $\hat{\textbf{p}}$, while maintaining a short codeword length.

\subsection{Adaptive Preprocessing}

\begin{figure}[htbp]
    %\vspace{-4mm}
	\centerline{
        \includegraphics[scale=0.53]{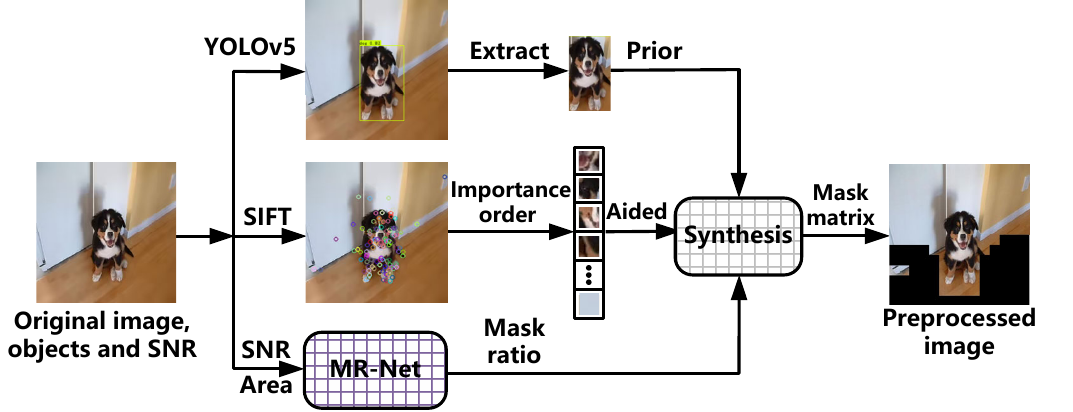}}
        \caption{The pipeline of the adaptive preprocessing module, consisting of three parallel processes: YOLOv5 for object extraction, SIFT for patch importance ordering, and MR-Net for determining the MR.} 
	\label{fig2}
    \vspace{-3.5mm}
\end{figure}

\begin{figure*}[!t]
    \vspace{-2mm}
	\centerline{\includegraphics[width=6.3in]{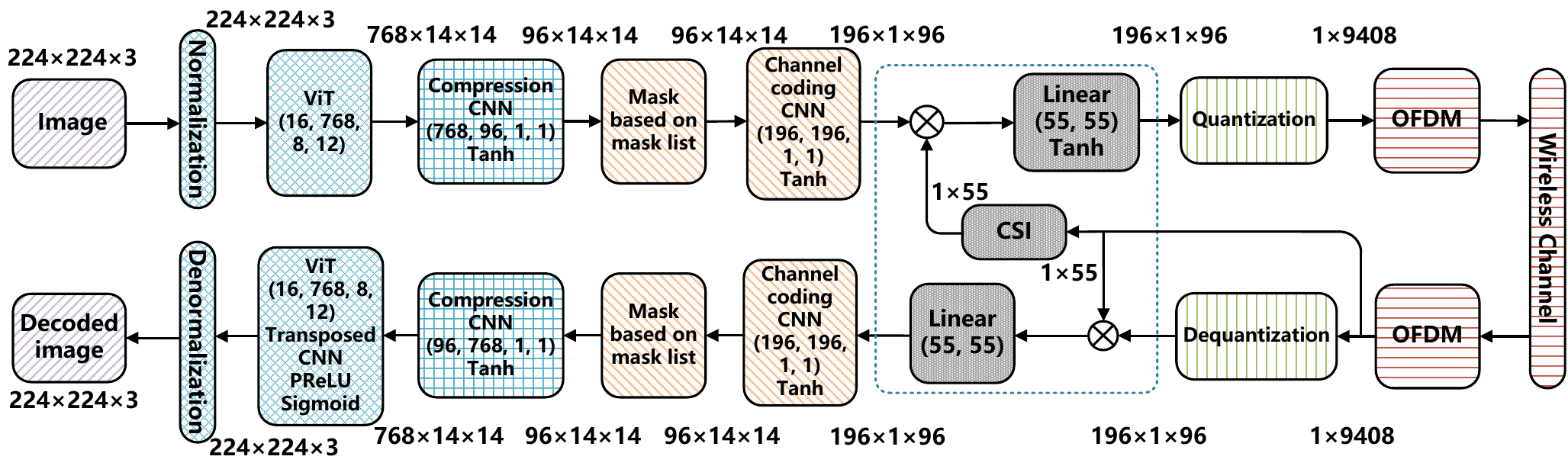}}
	\caption{The structure of the JSCC NN. For ${\rm ViT} (a, b, c, d)$ , the parameters $a$, $b$, $c$, and $d$ denote patch size, embedding dimension, head number, and block number, respectively. For ${\rm CNN} (a, b, c, d)$, $a$, $b$, $c$, and $d$ represent input channel, output channel, kernel size, and stride, respectively. The section framed by the dotted line is the optional CSIPA-Net. For ${\rm Linear} (a, b)$, $a$ and $b$ denote the input and output neuron numbers, respectively. The numbers next to each layer indicate the network output dimensions.}
	\label{fig3}
    \vspace{-5mm}
\end{figure*}

The pipeline for the adaptive preprocessing module is illustrated in Fig. \ref{fig2}. This module is designed for scenarios with fixed and limited bandwidth, where the number of symbols transmitted over the wireless channel is constant. To prioritize different parts of the image, the image is divided into multiple equal-sized patches, following the approach of vision transformers \cite{52}. The key concept is to mask certain patches while leveraging the redundancy of these masked patches to help recover the unmasked patches during transmission. The JSCC NNs facilitate this by ensuring that important patches are protected using the semantic coding scheme, as explained in the next subsection. Under varying wireless channel conditions, ASCViT-JSCC dynamically adjusts the mask ratio (MR) to balance the trade-off between information preservation and redundancy. A higher MR indicates that more patches in the image are masked.
 
The adaptive preprocessing module consists of three main steps:
\begin{enumerate} 
\item {\bf Object Detection with YOLOv5:} The YOLOv5 object detection algorithm is used to detect and extract objects in the image. For example, as shown in Fig. \ref{fig2}, the object ``dog'' is enclosed within a bounding box \cite{49}. Patches within this bounding box are given the highest priority, meaning they must be retained and protected.

\item {\bf Patch Importance Analysis with SIFT:}  The SIFT algorithm is employed to detect feature points within the image patches. The number of feature points in each patch indicates its importance---patches with more feature points are deemed more critical and are thus retained and protected with higher priority.

\item {\bf Importance Ordering:} Based on the object detection and the number of feature points in each patch, an importance order is generated. This order establishes the priority of each patch for encoding and transmission.
\end{enumerate}

Once the importance order is established, the module determines the appropriate MR using the MR-Net, as depicted in Fig. \ref{fig2}. MR-Net is a dense NN that calculates the MR based on two key inputs: the SNR of the channel and the area of the detected objects (i.e., the number of pixels occupied by all bounding boxes). When the SNR is high and the object area is small, MR-Net outputs a lower MR, ensuring that more informative patches are retained. Conversely, when the SNR is low or the object area is large, MR-Net outputs a higher MR, masking more patches as redundant to protect the unmasked patches.
 
With this dynamic adjustment, ASCViT-JSCC can generate an optimal MR based on channel conditions. By combining the MR and the importance order, ASCViT-JSCC generates an optimal mask list, represented as a binary sequence corresponding to the number of image patches. In this sequence, a value of 0 indicates that a patch is masked, while a value of 1 means the patch is retained. The final mask matrix $\textbf{m}$, consisting of binary values, is then applied to the input image to produce the preprocessed image $\textbf{p}$, which is ready for transmission.

\subsection{JSCC NN Structure}
To leverage the masked patches for recovering unmasked patches and encode the masked image $\textbf{p}$ into codewords, we propose a JSCC NN structure based on ViT and CNN, as shown in Fig. \ref{fig3}.

The input to the JSCC NN consists of the preprocessed image $\textbf{p}$ and the mask matrix $\textbf{m}$.  First, the NN normalizes the pixel values of the input image to the range $[0, 1]$ to facilitate training. The ViT is then used to extract semantic information from the normalized image, and a compression CNN maximally compresses the extracted information. These two processes are similar to the approaches used in the existing JSCC work \cite{48}. 

To train the network to treat masked patches as redundancy and use them to protect unmasked patches, we introduce a mask operation followed by a channel coding CNN that treats all patches as a single channel. This process is also applied at the decoding stage. Following this, a quantization module converts the floating-point data output from the networks into semantic modulation data. In our study, the quantized semantic modulation data is chosen from $[-3, -1, 1, 3]$, corresponding to 16-ary quadrature amplitude modulation (16-QAM). For higher modulation orders, such as 64-QAM, the data can be selected from $[-7, -5, -3, -1, 1, 3, 5, 7]$, though this would require retraining the network for the new modulation scheme. A higher modulation order increases quantization precision and accelerates convergence of the NNs. Note that the quantization and dequantization modules do not contain trainable parameters.

One challenge with quantization is that it leads to gradient truncation, making end-to-end training difficult. To overcome this, we adopt a simple quantization approach \cite{68} by passing gradients from the decoder to the encoder without modification, enabling efficient backpropagation during training. As a result, the last layer of the NN uses a sigmoid activation function to support quantization.

The signal processing at the physical layer and over the wireless channel is integrated into the JSCC NN framework. The receiver network mirrors the design of the transmitter network, and all network components are differentiable to allow for end-to-end optimization. At the transmitter, the JSCC NN outputs 9,408 baseband modulation symbols, resulting in a bandwidth compression ratio (BCR) of ${1}/{16}$ \cite{14}, which is fixed in our experiments.

During the training stage, only the ViT and compression CNN components are optimized, while other modules are excluded. The loss function used in this stage is the mean square error (MSE) between the preprocessed image $\textbf{p}$ and the decoded image $\hat{\textbf{p}}$, expressed as:  
\begin{equation}
MSE(\textbf{p}, \hat{\textbf{p}})=\frac{1}{H\times{W\times{C}}}\sum_{i=1}^{H\times{W\times{C}}}(p_i-\hat{p_i})^2, \label{eq6} 
\end{equation}
where $p_i$ and $\hat{p_i}$ represent the $i$-th pixel value of the preprocessed and decoded images, respectively. By minimizing the MSE, the ViT and compression CNN are optimized to efficiently perform pixel-level image reconstruction.

After this stage, the ViT and compression CNN parameters are frozen, and additional components such as the channel coding CNNs and quantization modules are added to the network. The full network is then trained in an additive white Gaussian noise (AWGN) channel, but only the channel coding CNNs are optimized during this phase. To simulate channel conditions, a random masking operation is used to instruct the channel coding CNN to prioritize the protection of unmasked patches. The loss function for this stage is 
\begin{equation}
MSE_{\rm{um}}(\textbf{p}, \hat{\textbf{p}})=\frac{1}{H\times{W\times{C}}}\sum_{i=1}^{H\times{W\times{C}}}(p_i-(\hat{p}_i \cdot m_i))^2\times{\frac{N_{\rm{T}}}{N_{\rm{U}}}}, \label{eq7} \end{equation}
where $N_{\rm{T}}$, $N_{\rm{U}}$ and $m_i$ represent the total number of patches, the number of unmasked patches, and the $i$-th value of the mask matrix, respectively. This function calculates the MSE for the unmasked patches, further training the network to prioritize their accurate reconstruction.

Finally, the parameters of all the NNs are fine-tuned using a small learning rate to achieve optimal performance under both AWGN and Rayleigh fading channels. For Rayleigh fading channels, least square (LS) channel estimation and zero-forcing (ZF) signal detection are incorporated to support the training process. The overall optimization goal is defined as 
\begin{equation}
\bm{\Theta^*}, \bm{\Phi^*} = \mathop{\arg\min}\limits_{\bm{\Theta},\bm{\Phi}}MSE_{\rm{um}}(\textbf{p}, \hat{\textbf{p}}),
\label{eq8} \end{equation}
where the parameters $\bm{\Theta^*}$ and $\bm{\Phi^*}$ represent the optimal parameters of the NNs that minimize the MSE between the unmasked patches of $\textbf{p}$ and $\hat{\textbf{p}}$. The masked patches are recovered using the MAE decoder.

\subsection{CSIPA-Net}
Most existing works on semantic communications focus primarily on AWGN and Rayleigh fading channels, as these can be easily integrated into NNs for end-to-end optimization. However, frequency-selective fading channels are more prevalent in real-world wireless communications. In this subsection, we address frequency-selective fading channels and demonstrate how they can be integrated into the NN framework to maintain differentiability. For simplicity, we assume that the channel gains for each subchannel in the OFDM system remain constant during the transmission of an image.

In conventional communication systems, transmitters often use CSI fed back from the receiver to enhance performance. Building on this approach, we assume that the receiver can estimate the CSI of all subchannels using LS channel estimation with the aid of comb pilots. Specifically, let there be $K$ subchannels, with each subchannel having a frequency response $h_j$, where $j$ is the index of the subchannel ($j=1, 2, \dots, K$). The frequency responses of all subchannels can be represented as $\textbf{h}=[h_1, h_2, \dots, h_{K}]$. The baseband modulation symbols $\textbf{x}$ are assigned to these $K$ subchannels, i.e., $\textbf{x}=[x_1, x_2, \dots, x_K]$. After transmission through the wireless channel, the received signal can be expressed as
\begin{equation}
\textbf{y}=\textbf{h} \odot \textbf{x}+\textbf{z},
\label{eq9}     
\end{equation}
where $\textbf{z}$ represents AWGN with mean $0$ and variance ${\sigma}^2$. Assuming that the receiver estimates the channel response as $\hat{\textbf{h}}=[\hat{h}_1,  \hat{h}_2, \dots, \hat{h}_K]$ , the estimated received codewords are given by $\hat{\textbf{x}}=[\frac{y_1}{\hat{h}_1}, \frac{y_2}{\hat{h}_2}, \dots, \frac{y_K}{\hat{h}_K}]$. The total SNR of the system is expressed as
\begin{equation} 
SNR=\frac{\sum_{k=1}^K  |h_k\cdot x_k|^2 }{K\sigma^2}.
\label{eq10}     
\end{equation}

To reduce feedback overhead and enhance system robustness, the receiver sorts the subchannels in descending order based on their energy. Rather than feeding back the full CSI, the receiver only sends the order of the subchannels to the transmitter. This approach simplifies transmitter design and reduces the accuracy requirements for channel estimation, improving overall system robustness.

To further optimize system performance by leveraging CSI, we propose a separate, optional module called CSIPA-Net. The position of CSIPA-Net within the system is illustrated in Fig. \ref{fig3}. Its architecture primarily consists of linear layers with Tanh activation functions. At the transmitter side, multiplication operations, a linear NN, and an activation function are applied before the quantization module. Similarly, these operations are performed after the dequantization module at the receiver side.

This design allows CSIPA-Net to optimize power allocation dynamically, learning to allocate more power to subchannels with higher SNRs. In essence, CSIPA-Net remaps the more crucial features to subchannels with better channel conditions, thereby enhancing the overall performance of the system. During training, all other network parameters are frozen, and CSIPA-Net is optimized independently. This process can be formulated as  
\begin{equation}
\bm{\Omega^*} = \mathop{\arg\min}\limits_{\bm{\Omega}}MSE_{\rm{um}}(\textbf{p}, \hat{\textbf{p}}),
\label{eq9} \end{equation}
where $\bm{\Omega}$ represents the parameters of CSIPA-Net, and $\bm{\Omega^*}$ are the optimal parameters that minimize the MSE between the unmasked patches of the transmitted and received images. 

%%%%%%%%%%%%%%%%  3 SIMULATION RESULTS %%%%%%%%%%%%%%%%%%%%%%
\section{Simulation Results}  
This section presents the numerical results and analysis of the proposed ASCViT-JSCC scheme. All NNs  were trained on an NVIDIA Tesla V100 GPU (32GB), while performance testing was conducted on an NVIDIA GTX 1650 GPU (4GB) to match the computational power of the proposed prototype validation platform.
 
\subsection{Configurations of the Simulation System}
\textbf{Datasets.} All experiments were conducted on the ImageNet2012 dataset \cite{56}. We randomly selected 20,000 images as the training set and 1,000 images each for the validation and test sets. All images were resized to $224\times{224\times{3}}$, and the patch size was set to $16\times16$, resulting in 196 patches per image.

\textbf{Simulation settings.} The SISO-OFDM system employed in this study consists of 64 subcarriers, with 55 allocated for data transmission and 9 for comb pilots. We used LS channel estimation, ZF  signal detection, and 16-QAM modulation. For object detection using YOLOv5, the intersection over union (IoU) threshold was set to 0.3, and the confidence threshold (CT) was set to 0.5.

\textbf{Training settings.}  The batch size was set to 8, and the Adam optimizer with a learning rate of 0.0002 was used for training. The networks were trained at an SNR of 10dB, as well as at SNRs uniformly sampled from the range of $[-5, 15]$ dB.

\textbf{Comparison schemes.} We considered two comparison schemes to evaluate the performance of the proposed ASCViT-JSCC:

1) \textbf{BPG, LDPC, and 16-QAM.} BPG (Better Portable Graphics) \cite{58} is an efficient image compression algorithm. We used LDPC coding with a code length of 1440 and a one-half code rate. The number of modulation symbols was controlled to be approximately equal to those in our approach. If BPG failed to decode the received bits, a grayscale image was generated, with the result set to 0.

2) \textbf{DeepJSCC with quantization.} We used DeepJSCC \cite{14} as another DL-based JSCC semantic transmission scheme, incorporating the same quantization modules as our proposed approach. 

\begin{figure} [htbp]
    \vspace{-5mm}
	\centering  
    \subfloat[PSNR+CS versus MR]{
        \centering  
        \includegraphics[scale=0.47]{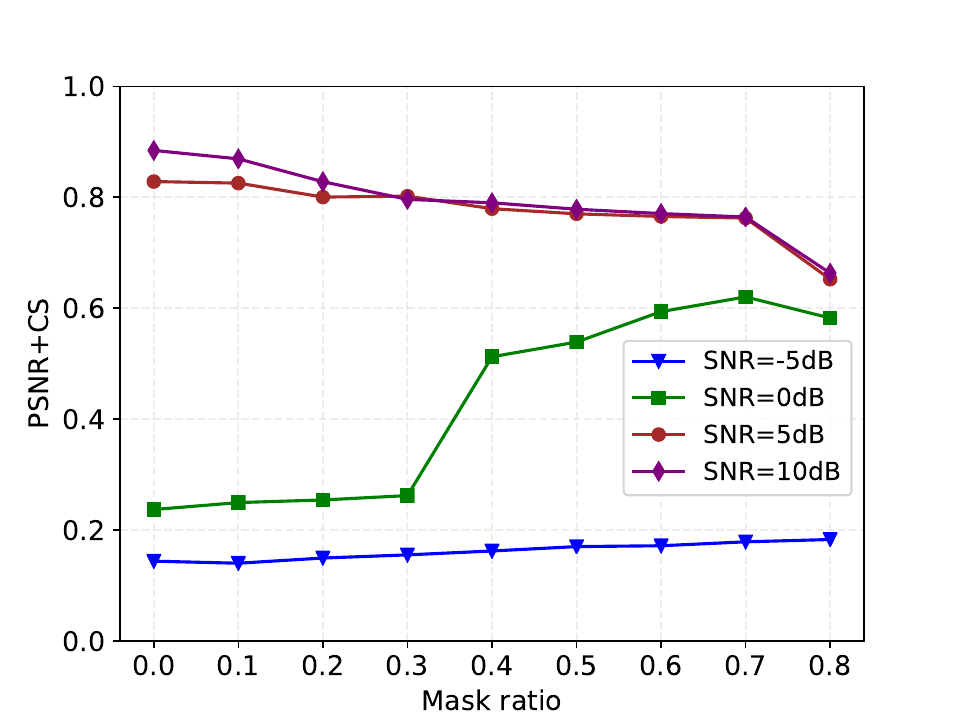} 
        \label{fig4:a}
    }
    \vspace{-4mm}
    \  
	\subfloat[SSIM+CS versus MR]{
        \centering  
        \includegraphics[scale=0.47]{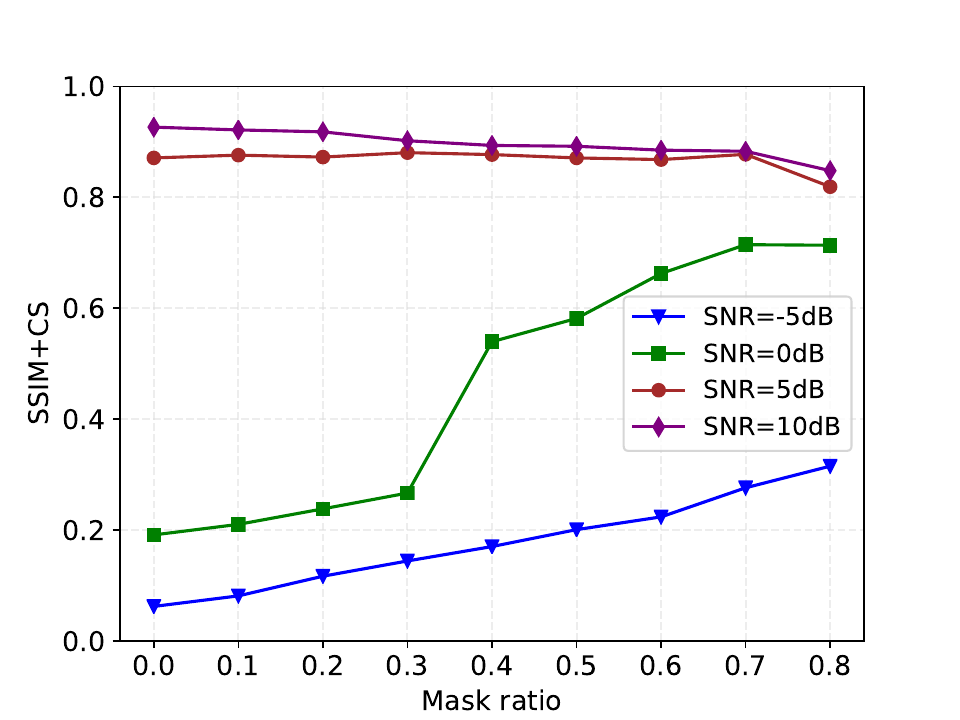} 
        \label{fig4:b}
    }
	\caption{Performance metrics versus MR. All results are measured in an AWGN channel.}
    \label{fig4}
    \vspace{-4mm}
\end{figure}

\begin{figure}[htbp]
    \vspace{-2mm}
	\centerline{
    \includegraphics[scale=0.48]{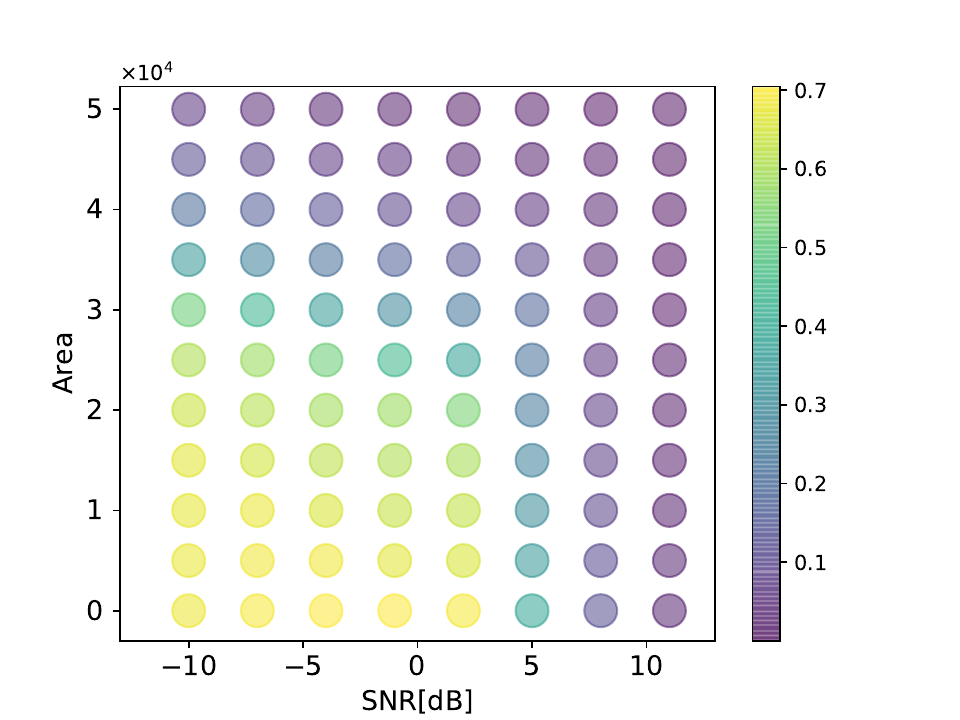}}
	\caption{MRs versus object areas and SNRs.} 
	\label{fig5}
    \vspace{-4mm}
\end{figure}

\textbf{Metrics.} We used three independent metrics and combined them into two comprehensive metrics \cite{59}:

1) \textbf{Peak Signal-to-Noise Ratio (PSNR).} PSNR measures the MSE between two images and considers the image depth. It is calculated as
\begin{equation}
PSNR=10 \log_{10}\left(\frac{MAX^2}{MSE}\right),
\label{eq12}
\end{equation}
where $MAX$  is the maximum pixel value, and $MSE$  is the mean squared error between the images, similar to (\ref{eq6}).

2) \textbf{Structural Similarity Index Measure (SSIM).} SSIM assesses the structural similarity between two images using a sliding window. Given windows $x$ and $y$ in two images, SSIM is calculated as
\begin{equation}
SSIM=\frac{(2\mu_x \mu_y+c_1)(2\sigma_{xy}+c_2)}{(\mu_x^2+\mu_y^2+c_1)(\sigma_x^2+\sigma_y^2+c_2)}
\label{eq13},
\end{equation}
where $\mu_x$, $\mu_y$, $\sigma_x$, $\sigma_y$ and $\sigma_{xy}$ denote the means, variances, and covariance of $x$, $y$, respectively. Constant $c_1$ and $c_2$ avoid division by zero.

3) \textbf{Confidence Score (CS).} This metric, derived from YOLOv5, represents the quality of the detected objects in an image. If multiple objects are present, the average of all object CSs is used.

4) \textbf{Comprehensive metrics.} We combine the three independent metrics into two comprehensive metrics: PSNR+CS and SSIM+CS. PSNR is normalized to  $(0,1]$ using its maximum value $\delta$, and the combined metric is calculated as
\begin{equation} 
PSNR+CS=\frac{PSNR}{\delta}\times{\frac{1}{2}}+CS\times\frac{1}{2}. \label{eq14}
\end{equation}
Similarly, SSIM+CS is calculated as 
\begin{equation} 
SSIM+CS=SSIM\times{\frac{1}{2}}+CS\times\frac{1}{2}. \label{eq15} 
\end{equation}
The equal weighting 1/2 reflects the importance of both object quality and image reconstruction.
In AWGN channels, we report the results in terms of the three independent metrics, while other scenarios are evaluated using the two comprehensive metrics.

\begin{figure*}[htbp]
    \hspace{7mm}
	\subfloat[AWGN channel]{
		\includegraphics[width=2.6in]{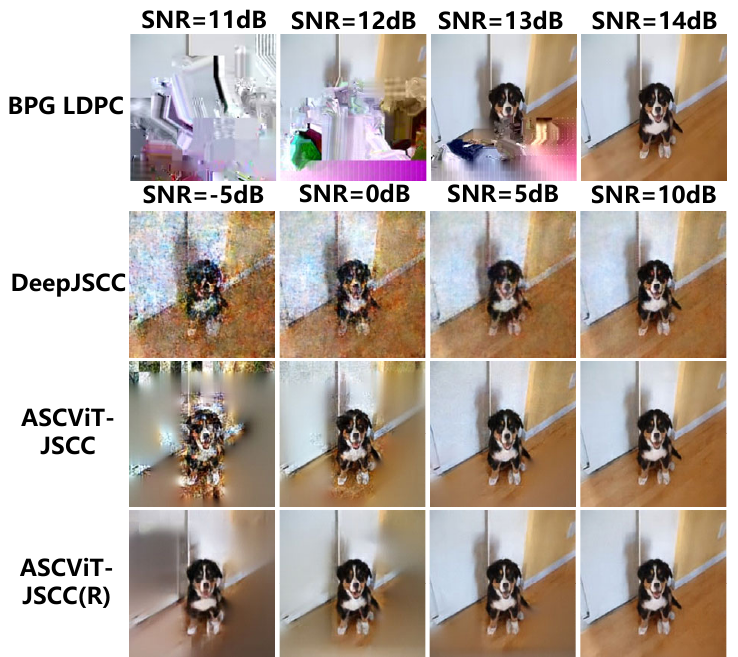}
        \label{fig6:a}
    }
    \hspace{20mm}
    \subfloat[Rayleigh fading channel]{
		\includegraphics[width=2.6in]{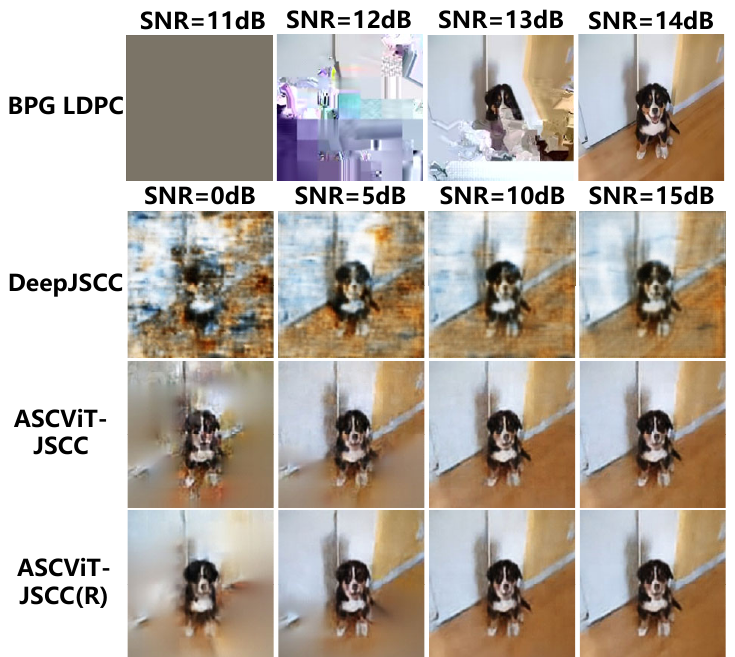}
        \label{fig6:b}
    }
        \caption{Visual results. ``R'' indicates that the network is trained with SNRs uniformly sampled from the range $[-5, 15]$dB, while other networks are trained at 10 dB. At the same SNR, ASCViT-JSCC and ASCViT-JSCC(R) utilize the same MR.}
	\label{fig6}
    \vspace{-5mm}
\end{figure*}

\subsection{Performance versus MR}
To evaluate performance versus MR and identify the optimal MRs at different SNRs, we use the network trained at 10 dB in the AWGN channel. Note that the number of baseband modulation symbols is constant at 9,408 and only the MR changes.

As shown in Fig. \ref{fig4}, both PSNR+CS and SSIM+CS exhibit substantial variations across different MRs at 0 dB SNR, while the metrics change more gradually under other conditions. From Fig. \ref{fig4}(a), we observe that PSNR+CS decreases as the MR increases at 5 dB and 10 dB SNRs. This suggests that the reconstructed image quality is already sufficiently high under these conditions, making additional masking unnecessary. Conversely, at 0 dB SNR, PSNR+CS improves as the MR increases, particularly in the range from 0.3 to 0.4. However, when the MR increases from 0.7 to 0.8, PSNR+CS decreases, indicating that a MR of 0.7 is optimal for this SNR. Excessive masking at higher ratios could result in masking patches that are important for reconstruction. At lower SNRs, such as -5 dB, PSNR+CS consistently improves with increasing MRs, due to significant enhancements in object quality. The trend for SSIM+CS follows a similar pattern as PSNR+CS, although SSIM+CS shows less variation at higher SNRs and more pronounced changes at lower SNRs as MR varies. This suggests that the structural quality of reconstructed images is more sensitive to masking at lower SNRs.
 
Based on these observations, we can determine the optimal MRs for different SNRs. For example, at -5 dB SNR, a MR of 0.8 provides the best performance, while at 0 dB SNR, a MR of 0.7 is optimal. To avoid masking important objects and ensure better performance, we restrict the MR range to $[0, 0.7]$, preventing MR-Net from masking critical areas of the image. 

We trained MR-Net using a dataset constructed from the results above. Fig. \ref{fig5} shows the MRs versus object areas and SNRs. It is clear that the MR decreases as both the SNR and object size increase. This behavior is consistent with expectations, as larger objects and higher SNRs require fewer masked patches for effective reconstruction. These results highlight the adaptability of MR-Net to varying channel conditions and user requirements. With this refined MR-Net structure and parameter design, ASCViT-JSCC can achieve optimal performance. In subsequent experiments, the structure and parameters of MR-Net are fixed to ensure stable system performance. 

\subsection{Performance of ASCViT-JSCC in AWGN Channel}

This subsection compares the performance of different algorithms in AWGN channels. The visual results are shown in Fig. \ref{fig6}(a). It is evident that the image quality in traditional schemes changes drastically over a narrow SNR range. With just a 3 dB difference, the quality of images can transition from poor to excellent, illustrating the well-known \textit{cliff effect}\footnote[1]{The performance sharply declines when the channel capacity falls below the communication rate.} associated with traditional schemes. In contrast, NN-based schemes effectively mitigate this issue. Additionally, ASCViT-JSCC, which was trained at a fixed 10 dB SNR, outperforms DeepJSCC across all tested SNR levels. Furthermore, ASCViT-JSCC trained on random SNRs demonstrates superior object quality in low-SNR regimes compared to the model trained at a fixed SNR, while both exhibit similar performance in high-SNR regimes.

\begin{figure} [htbp]
    %\vspace{-3mm}
	\centering  
    \subfloat[PSNR versus SNR]{
        \centering  
        \includegraphics[scale=0.55]{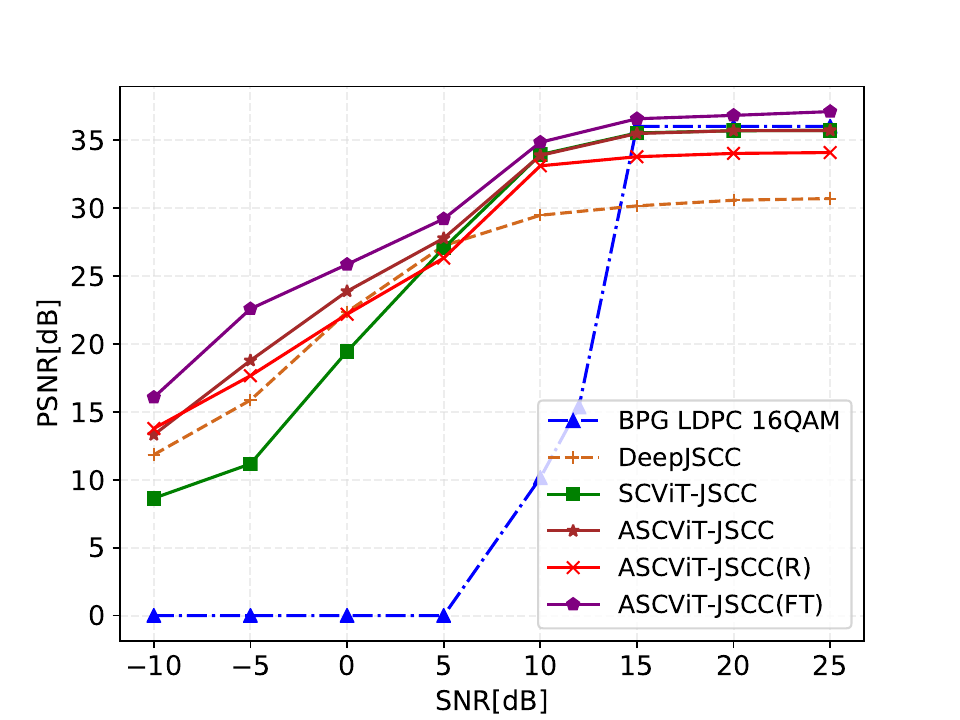} 
        \label{fig7:a}
    }
    \vspace{-4mm}
	\subfloat[SSIM versus SNR]{
        \centering  
        \includegraphics[scale=0.55]{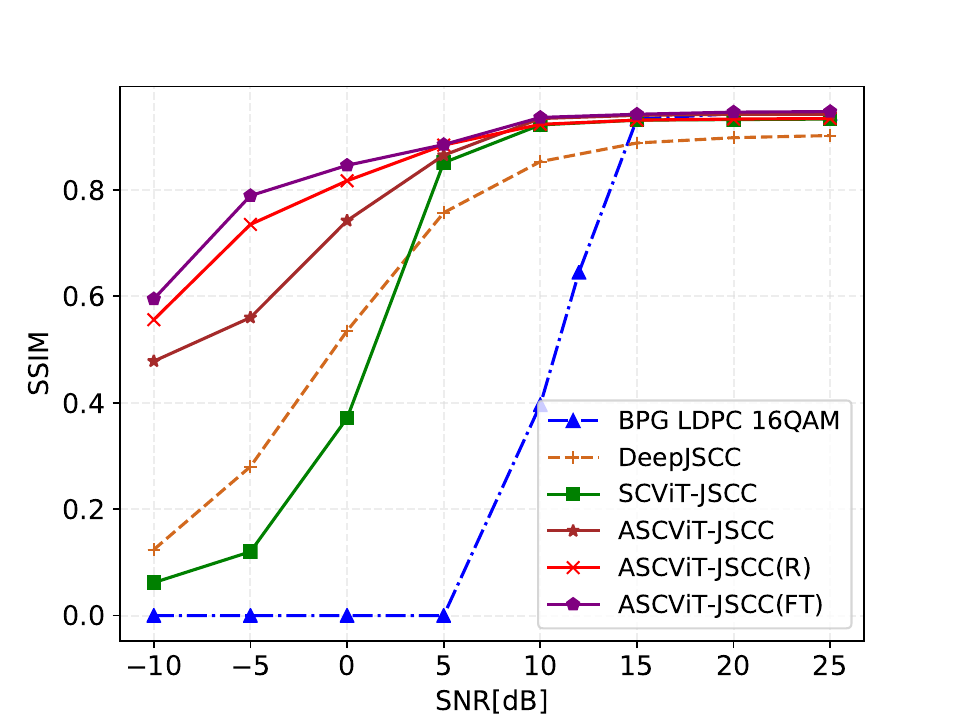} 
        \label{fig7:b}
    }
    \vspace{-4mm}
    
    \subfloat[CS versus SNR]{
        \centering  
        \includegraphics[scale=0.55]{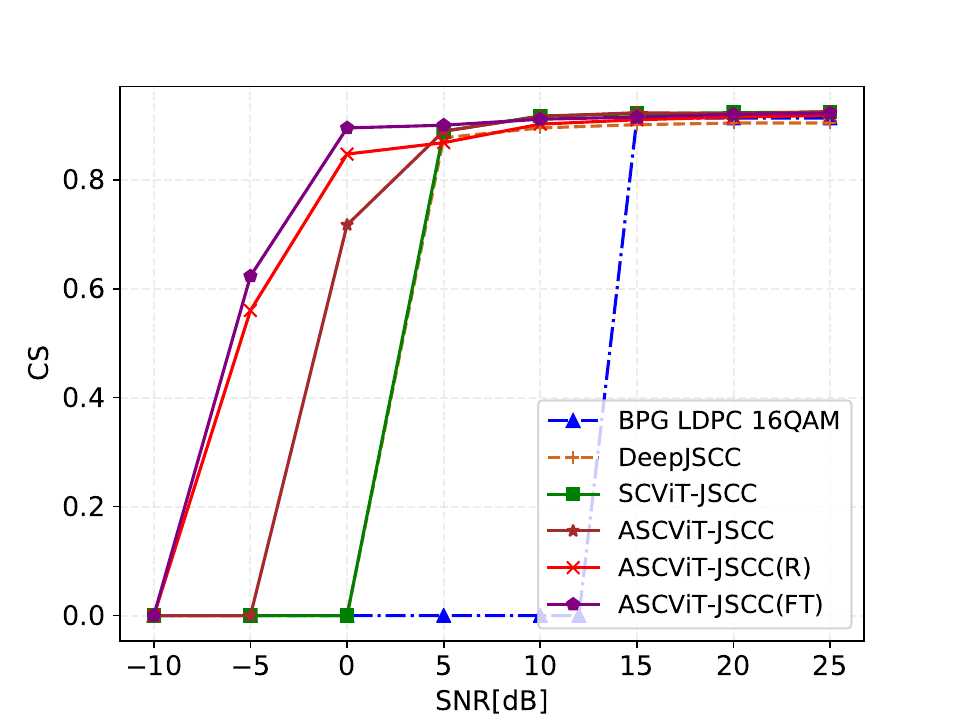} 
        \label{fig7:c}
    } 
    \caption{Performance of ASCViT-JSCC compared to other schemes in AWGN channels. DeepJSCC refers to the original DeepJSCC with quantization modules. SCViT-JSCC refers to ASCViT-JSCC without adaptive preprocessing. ``R" and ``FT" indicate that the NNs are trained at random SNRs uniformly sampled from [-5, 15] and fine-tuned at the tested SNR values, respectively.}

    \label{fig7}
    %\vspace{-5mm}
\end{figure}

Fig. \ref{fig7} presents numerical results at various SNRs in AWGN channels. The traditional scheme's performance improves sharply as the SNR increases from 10 dB to 15 dB, whereas NN-based schemes show smoother performance curves. ViT-based SCViT-JSCC outperforms CNN-based DeepJSCC in high-SNR regimes, while DeepJSCC performs better in low-SNR conditions due to the stronger reliance on distinct patch contours. In low-SNR regimes, ASCViT-JSCC outperforms SCViT-JSCC, while both schemes perform similarly in high-SNR conditions, indicating the effectiveness of the adaptive masking operation. As channel conditions improve, MR-Net outputs a MR of 0, causing ASCViT-JSCC and SCViT-JSCC to achieve the same performance.

NNs trained at random SNRs outperform those trained at fixed SNRs in terms of SSIM and CS in low-SNR regimes, but they perform worse in terms of PSNR. This is because networks exposed to harsher channel conditions during training tend to produce better object quality under low-SNR conditions in the testing phase, but this focus on object quality comes at the cost of lower PSNR.
 
Overall, our proposed scheme outperforms DeepJSCC in all SNR regimes and matches the performance of traditional schemes even in high-SNR conditions. ASCViT-JSCC, like other NN-based schemes, mitigates the \textit{cliff effect}, improving image quality in poor channel conditions. Additionally, due to the use of SIFT, the algorithm remains effective in extracting key patches and protecting them even in cases where YOLOv5 cannot detect objects in the images.

\subsection{Performance of ASCViT-JSCC in Rayleigh Fading Channel}

Rayleigh fading channels represent a critical scenario in wireless communications, given their relevance to environments with no line-of-sight. In this context, we fine-tuned the NN parameters using models trained at 10 dB SNR in the AWGN channel. Visual results from the four schemes are presented in Fig. \ref{fig6}(b), and it is clear that overall image quality degrades in Rayleigh channels compared to AWGN channels. Traditional schemes still exhibit significant quality variation across a narrow 4 dB range. For NN-based schemes, image quality is notably reduced, yet ASCViT-JSCC continues to outperform DeepJSCC, highlighting the benefits of adaptive masking.

\begin{figure*} [htbp]
    \hspace{-2mm}
	\centering
	\subfloat[PSNR+CS versus SNR]{
		\includegraphics[width=3.1in]{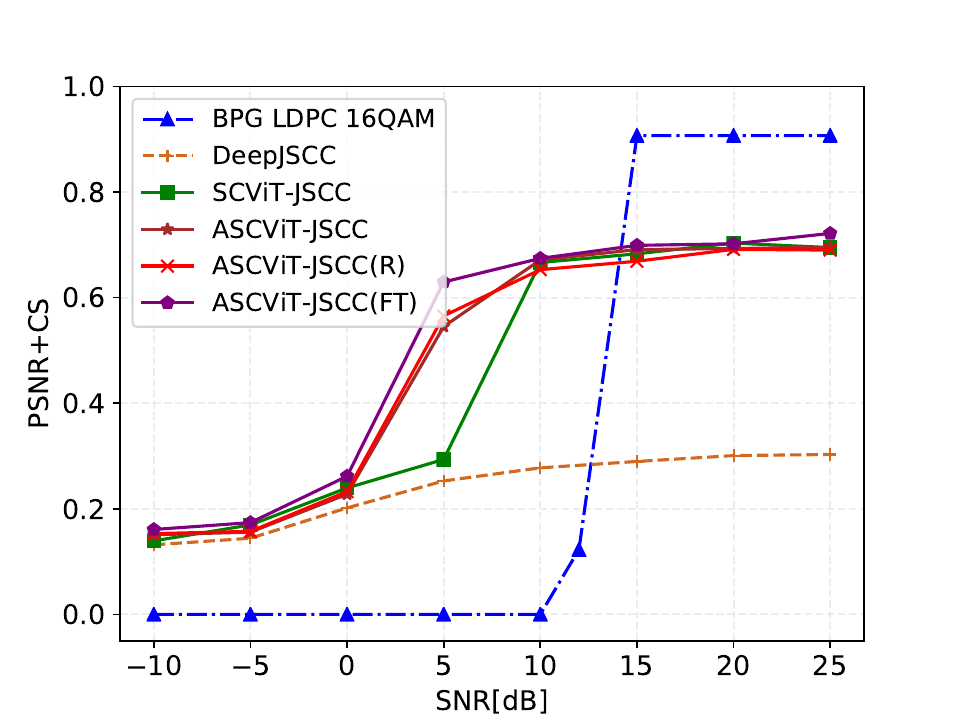}
        \label{fig8:a}
    }
	\subfloat[SSIM+CS versus SNR]{
		\includegraphics[width=3.1in]{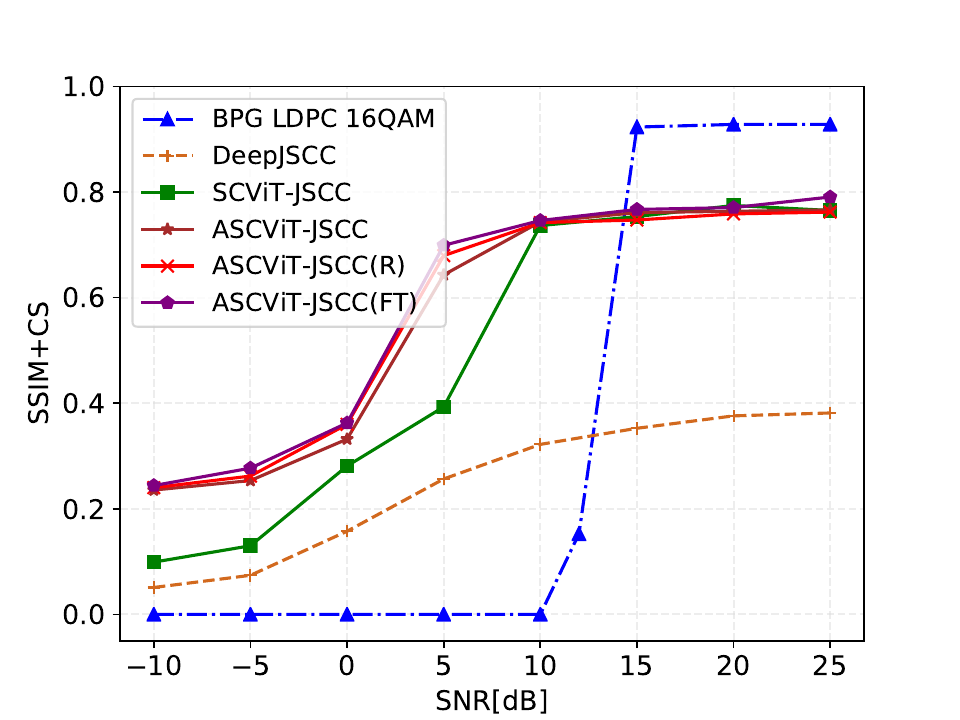}
        \label{fig8:b}
    }
	\caption{Performance of ASCViT-JSCC compared to traditional scheme, DeepJSCC, and SCViT-JSCC in Rayleigh fading channel. ``R" and ``FT" indicate that NNs are trained at random SNRs uniformly sampled from [-5, 15], and fine-tuned at the tested SNR values, respectively.}
	\label{fig8} 
    \vspace{-5mm}
\end{figure*}

Fig. \ref{fig8} provides the numerical results in Rayleigh fading channels. The performance curve for traditional schemes shifts to the right compared to AWGN channels, indicating that more SNR is required to achieve similar image quality. In terms of PSNR+CS, ASCViT-JSCC continues to outperform both SCViT-JSCC and DeepJSCC, but the adaptive masking operation shows its advantage only within a narrow SNR range. This suggests that the impact of adaptive masking becomes less significant in Rayleigh fading channels, likely due to the increased complexity of training  NNs in such environments and the direct use of MR-Net parameters optimized for AWGN channels.

An interesting observation is that ASCViT-JSCC trained with fixed and random SNRs performs similarly in Rayleigh channels. This might be due to the challenges of training NNs effectively in Rayleigh fading environments. Similarly, fine-tuning the networks does not result in significant performance improvements in this context. Furthermore, DeepJSCC performs substantially worse in Rayleigh channels compared to AWGN conditions. This drop in performance is primarily attributed to YOLOv5’s inconsistent object detection under severe fading, which negatively impacts the CS evaluation.

The trend in SSIM+CS mirrors that of PSNR+CS, although ASCViT-JSCC’s advantage over SCViT-JSCC becomes more pronounced in low SNR regimes. The adaptive preprocessing module appears to better handle the degradation caused by fading, underscoring the benefit of such mechanisms in weaker channel conditions. Despite the limitations of NN-based schemes in high SNR regimes under Rayleigh fading, their ability to mitigate the \textit{cliff effect} remains evident, particularly in low SNR conditions where traditional schemes struggle to maintain performance.

\subsection{Performance of ASCViT-JSCC with CSIPA-Net in Frequency-selective Channel}

\begin{figure}[t]
    \vspace{-3mm}
	\centerline{\includegraphics[width=3.1in]{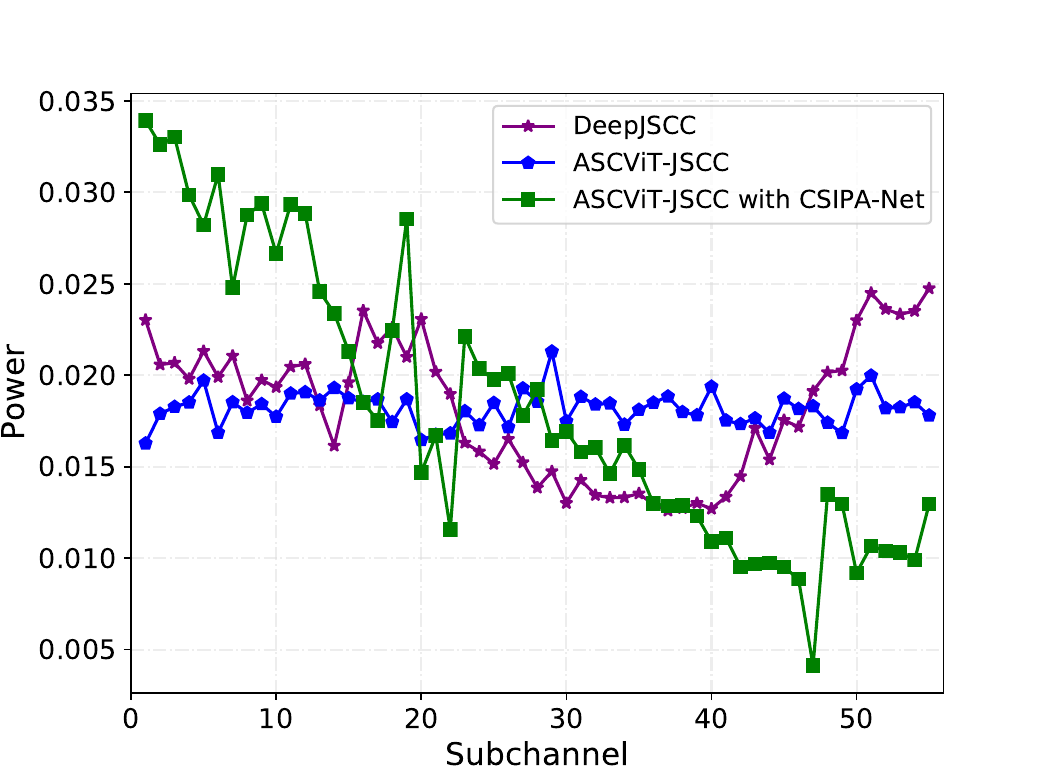}}
	\caption{Power allocation on each subchannel.}
	\label{fig9}
    \vspace{-5mm}
\end{figure}

In wireless communications, frequency-selective fading occurs when the signal bandwidth exceeds the channel coherence bandwidth. This subsection evaluates the performance of NN-based schemes with CSIPA-Net under frequency-selective fading conditions. The multipath channel used during training consists of three paths with an exponential power delay profile, and each path follows a complex Gaussian distribution. ASCViT-JSCC with CSIPA-Net is trained under these channel conditions, with perfect channel estimation and ZF  signal detection.

Fig. \ref{fig9} shows the power allocation on each subchannel. Both DeepJSCC and ASCViT-JSCC allocate power uniformly across subchannels, as these methods rely only on source distribution and system SNR, without leveraging CSI. In contrast, CSIPA-Net dynamically reallocates more power to subchannels with higher SNRs (those with greater energy), improving the transmission of critical features.

\begin{figure} [t]
    \vspace{-4mm}
	\centering
	\subfloat[PSNR+CS versus SNR]{
		\includegraphics[width=3.2in]{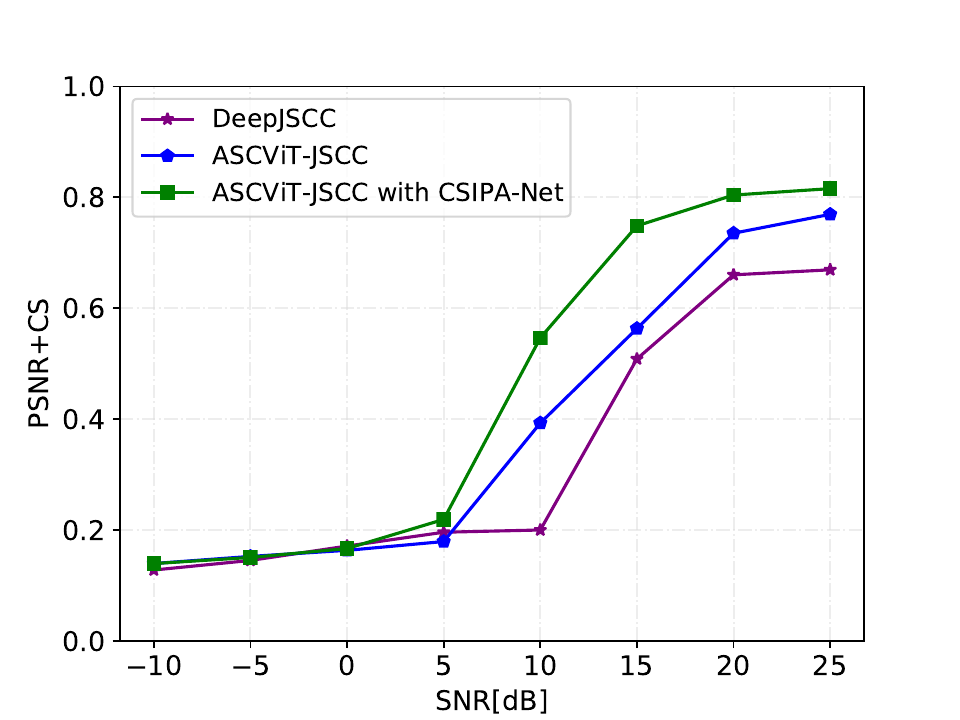}
        \label{fig10:a}
  }
  \vspace{-2mm}
  \
	\subfloat[SSIM+CS versus SNR]{
		\includegraphics[width=3.2in]{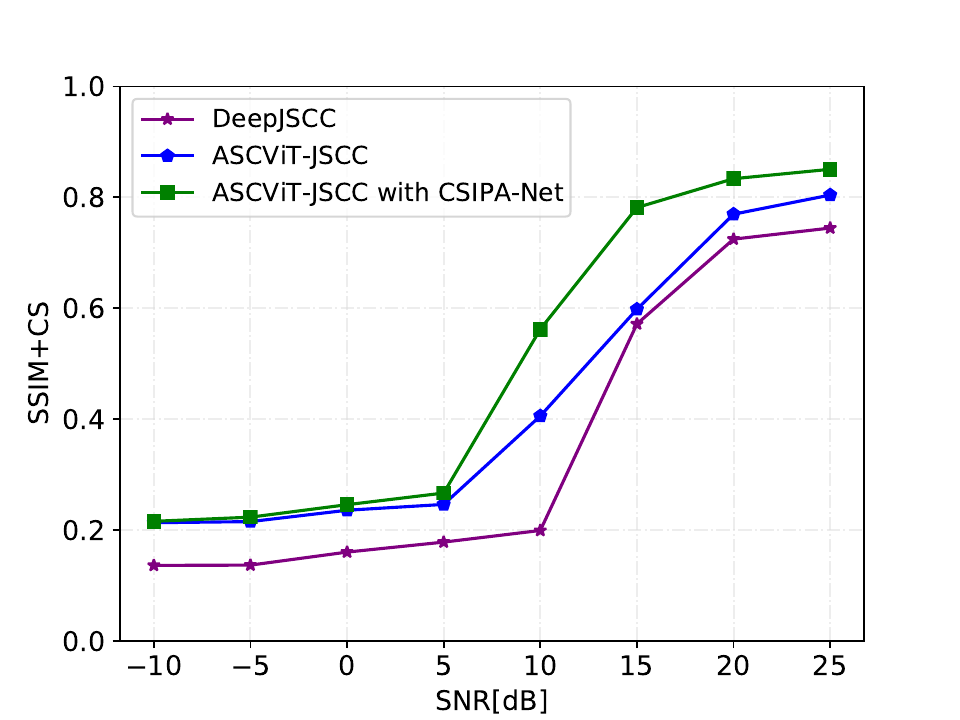}
        \label{fig10:b}
  }
	\caption{Performance of DeepJSCC, ASCViT-JSCC, and ASCViT-JSCC with CSIPA-Net in multipath channels with 5 paths.} 
	\label{fig10} 
    \vspace{-4mm}
\end{figure}

%\begin{figure}[htbp]
%	\centering
%	\includegraphics[width=3.1in]{figure/Raccoon.pdf}
%	\caption{Robustness testing of ASCViT-JSCC.}
%	\label{fig11} 
 %   \vspace{-5mm}
%\end{figure}

To further demonstrate the feasibility and robustness of CSIPA-Net, we evaluate the performance of three schemes---DeepJSCC, ASCViT-JSCC, and ASCViT-JSCC with CSIPA-Net---in a multipath channel with 5 paths, as shown in Fig. \ref{fig10}. During the evaluation, we manually distribute features to different subchannels according to their priorities. In Fig. \ref{fig10}(a), ASCViT-JSCC with CSIPA-Net outperforms ASCViT-JSCC at SNRs above 5 dB, while DeepJSCC consistently performs the worst. In the low SNR regime, all schemes struggle to recover images effectively due to poor channel conditions. As SNR increases and exceeds 25 dB, the performance gap between ASCViT-JSCC and ASCViT-JSCC with CSIPA-Net narrows and stabilizes. The trends in SSIM+CS, shown in Fig. \ref{fig10}(b), mirror those in PSNR+CS, further illustrating the advantage of CSIPA-Net.

% These results highlight the benefits of using CSIPA-Net for adaptive power allocation, particularly in frequency-selective channels. CSIPA-Net enables ASCViT-JSCC to allocate power intelligently, improving performance in challenging channel conditions. Additionally, the robustness of CSIPA-Net is evident, as it continues to perform well even when tested in channel conditions that differ from those encountered during training.

In summary, CSIPA-Net successfully enhances ASCViT-JSCC by rationally allocating power across subchannels, improving overall performance in frequency-selective channels.

%\subsection{Robustness Testing of ASCViT-JSCC}
%Finally, we evaluate the robustness of the proposed algorithm. ASCViT-JSCC leverages YOLOv5 to detect objects in images and prioritize preserving these parts. However, the number of categories that can be detected by YOLOv5 is limited by training dataset. Therefore, it is necessary to test the effectiveness of the proposed scheme for categories not included in training dataset. Specifically, we choose “raccoon” which is not included in the COCO dataset for testing. The results, shown in Fig. \ref{fig11}, indicate that even when YOLOv5 fails to work, the masking effect remains effective due to the presence of SIFT. 

%%%%%%%%%%%%%%%%  4 PROTOTYPE VALIDATION %%%%%%%%%%%%%%%%%%%%%%
\section{Prototype Validation}
The practical feasibility of semantic communications needs urgent validation. To address this, we developed a OFDM prototype validation platform called ICP, based on SDR and embedded GPU systems. This section outlines the system framework, hardware components, software design, experimental scenarios, deployment procedure, experimental results, performance analysis, and complexity assessment.

\subsection{System Framework and Hardware Components}
\begin{figure*}[htb]
    % \centering
    \hspace{1mm}
	\subfloat[System framework]{
        %\centering
		\includegraphics[width=3.5in]{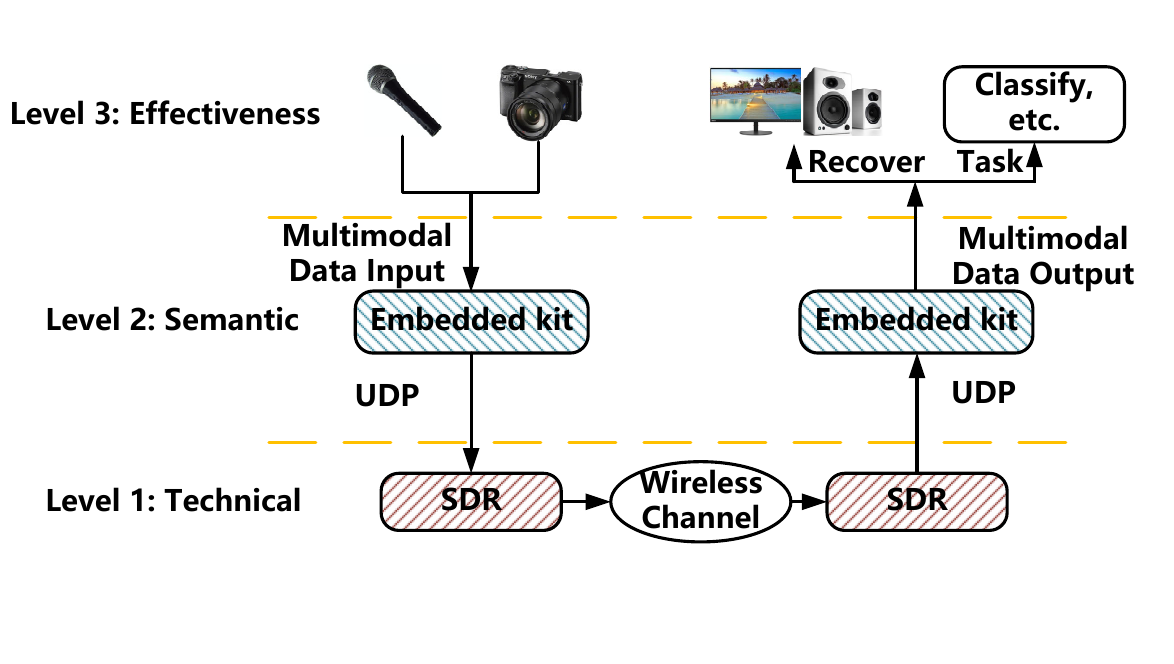}
        \label{fig11:a}
    }
    \hspace{4mm}
    \subfloat[Hardware components]{
        %\centering
		\includegraphics[width=3.3in]
        {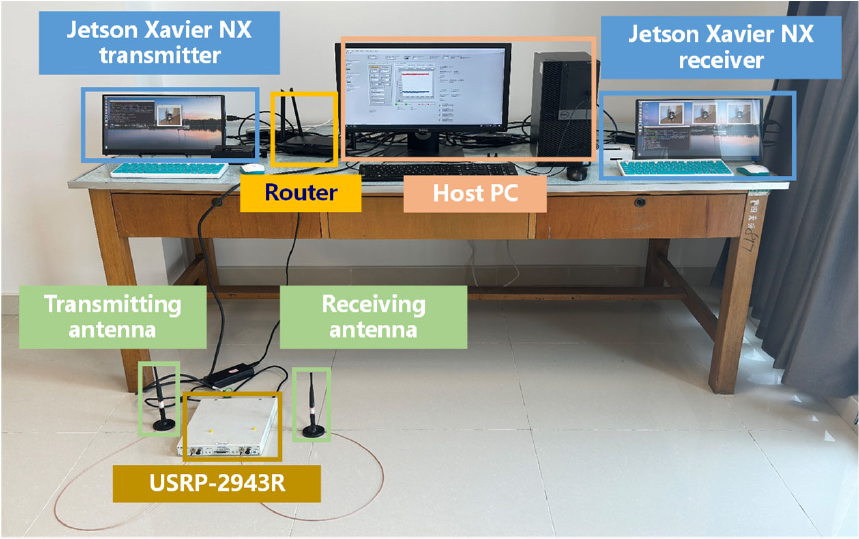}
        \label{fig11:b}
    }
    \caption{System framework and hardware components of the ICP.} 
	\label{fig11}
    \vspace{-2mm}
\end{figure*}

The system framework of the ICP is illustrated in Fig. \ref{fig11}(a). Similar to the communication hierarchy described in \cite{3}, the ICP is structured into three levels:
\begin{itemize}
    \item {\bf Effectiveness level:} Handles multimedia data acquisition, source recovery, and task execution.
    \item {\bf Semantic level:} Facilitates efficient source and channel codecs.
    \item {\bf Technical level:} Manages physical layer signal processing \cite{3}.
\end{itemize}
The effectiveness and semantic levels utilize common embedded communication protocols to exchange data. For communication between the semantic and technical levels, the UDP protocol is chosen for its ability to handle large data transmissions efficiently.
 
\begin{figure} [htbp]
	\centering
	\includegraphics[width=3.3in]{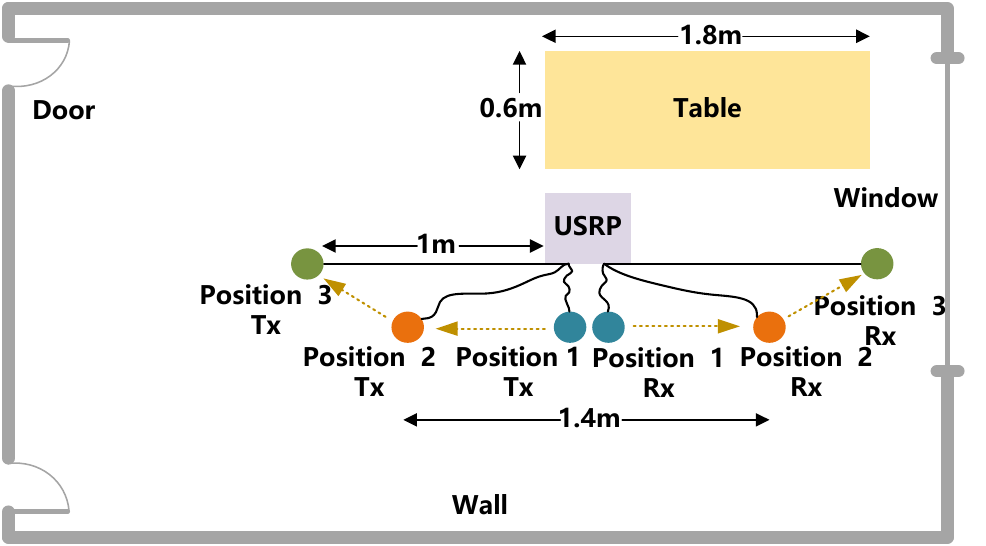}
	\caption{Testing scenarios. All devices are placed in the corner of a large office. The distance between two antennas at position 1 is 20 cm. The dotted lines indicate antenna movements.}
	\label{fig12}
    \vspace{-3mm}
\end{figure}

As shown in Fig. \ref{fig11}(b), the ICP consists of the following hardware components:
\begin{itemize}
    \item Two {\bf NVIDIA Jetson Xavier NX} devices, each featuring 384 NVIDIA CUDA cores and 48 Tensor cores, providing sufficient computational power for deploying DL models. These devices run on a Linux-based OS, supporting the creation of DL environments compatible with the ARM64 architecture.
    \item {\bf USRP-2943R} with two antennas, offering two RF transceivers with a bandwidth of 120 MHz. In this setup, one RF channel is used for transmission and the other for reception, allowing self-transmission and self-reception on the same device.
    \item A {\bf host PC} and {\bf router}, used to establish an Ethernet connection for UDP transmission and reception between the Jetson devices and the USRP-2943R.
    \item {\bf Antenna feeders} (1 meter long) for modifying channel environments to test different scenarios.
\end{itemize}

Compared to existing ICPs, the ICP is compact and supports flexible updates with commercial off-the-shelf components, facilitating the deployment of intelligent algorithms through high-level programming languages like Python.
 
\begin{figure*} [htbp]
	\centering
	\subfloat[PSNR+CS versus SNR]{
		\includegraphics[width=3.3in]{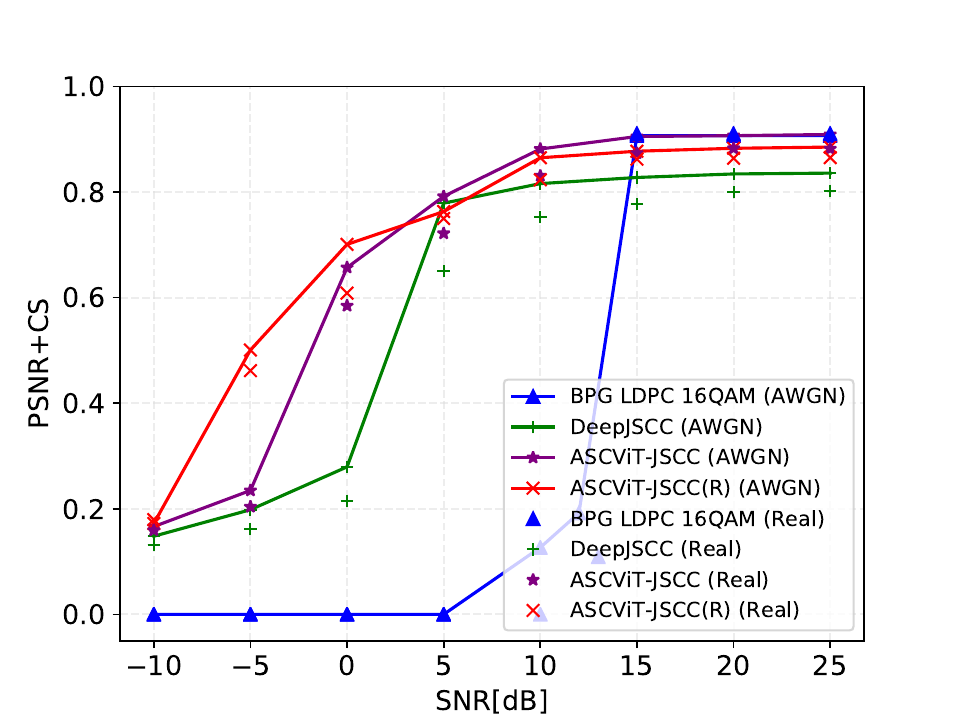}
        \label{fig13:a}
  }
	\subfloat[SSIM+CS versus SNR]{
		\includegraphics[width=3.3in]{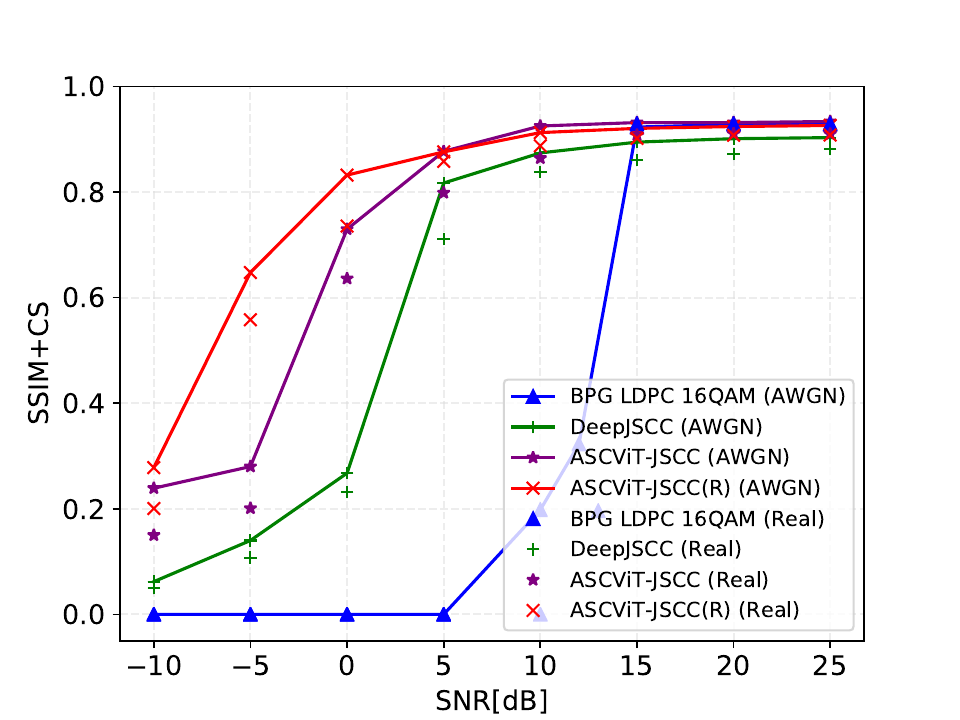}
        \label{fig13:b}
  }
	\caption{Comparison between performance measured in real channels by controlling noise manually and that in simulated AWGN channels. ``R" indicates that NNs are trained at ramdom SNRs uniformly sampled from [-5, 15]. ``AWGN" indicates the  results are from AWGN channel simulations and ``Real" indicates the results are measured in practise.}
	\label{fig13}
\end{figure*}

\subsection{Software Design}
\begin{table}[t]
  \renewcommand{\arraystretch}{1.5}
  \begin{center}
    \caption{Parameters of the ICP}
    \label{tab1}
    %\resizebox{1\columnwidth}{!}{
        \begin{tabular}{|c|c|c|c|} 
          \hline
          \textbf{Carrier frequency} & 2 GHz & \textbf{System bandwidth} & 0.364 MHz \\
          \hline
          \textbf{Sampling frequency} & 1 MHz & \textbf{Subcarrier spacing} & 3.906 kHz \\
          \hline     
          \textbf{Symbols per frame} & 41 & \textbf{FFT size} & 256 \\
          \hline
          \textbf{\makecell[c]{OFDM\\symbol duration}} & 0.32 ms & \textbf{Frame duration} & 13.12 ms \\
          \hline
        \end{tabular}
    %}
  \end{center}
  \vspace{-3mm}
\end{table}

The programming languages utilized in the ICP are primarily {\bf LabVIEW} and {\bf Python}. Python is used on the Jetson Xavier NX devices to implement NN algorithms, while LabVIEW is employed on the USRP-2943R to perform physical layer signal processing.
 
At the {\bf Jetson Xavier NX transmitter}, multimedia data is processed using Python, where it undergoes source and channel coding to produce bitstreams or modulation symbols. These data are transmitted to the USRP-2943R via the UDP protocol. The USRP-2943R then transforms the data into complex modulation symbols. Each group of 125 symbols is augmented with 25 pilot symbols, resulting in 150 symbols. To prevent interference with adjacent channels, 106 subcarriers are added as guard bands, bringing the total number of subcarriers in the frequency domain to 256. An inverse Fourier transform is then applied to generate an OFDM symbol, and a cyclic prefix (CP) of length 64 is added to mitigate inter-symbol interference. This results in an OFDM symbol with 320 samples.

Given the performance constraints of the USRP-2943R, the sampling rate is set to 1 million samples per second, and the carrier frequency is set to 2 GHz. Each frame contains 41 OFDM symbols and 600 synchronization samples, including 1 header symbol and 40 data symbols. This results in a frame duration of approximately 13.12 ms, and an effective data transmission rate of around 1.4 Mbps. The main parameters of the ICP are detailed in Table \ref{tab1}.
 
At the {\bf receiver}, the USRP-2943R samples incoming signals and converts them into digital form. After synchronization and frame extraction, 40 OFDM data symbols are processed through channel estimation and related techniques to reconstruct the bitstreams. These bitstreams are then sent to the Jetson Xavier NX receiver via UDP, where Python decodes them, reconstructs the original multimedia content, and displays it.

\subsection{Scenario Description and Deployment}
In this subsection, all schemes are deployed on the ICP platform simultaneously. Given the BCR  of $\frac{1}{16}$, each image requires only two frames for transmission using the designed frame structure.
 
To assess performance, we vary the channel environments. However, as channel conditions deteriorate, the SNR fluctuates significantly, making precise measurements challenging. Additionally, due to the limited mobility of the ICP and the 1-meter antenna feeder, only relatively simple scenarios can be tested. Therefore, we evaluate performance in two controlled scenarios:
\begin{enumerate}
    \item {\bf Scenario 1:} Noise power is manually controlled to ensure accurate and consistent results. To maintain reliable transmission, high transmission power and close proximity between the antennas, as shown in position 1 of Fig. \ref{fig12}, are used. The distance between antennas and transmission power remain fixed, while noise power is adjusted manually to simulate different SNR levels.
    \item {\bf Scenario 2:} The transmission power is fixed, and SNR is adjusted by varying the distance between the two antennas. As illustrated in Fig. \ref{fig12}, tests are conducted at three distances: 20 cm, 1.4 m, and 2 m, representing high, medium, and low SNR environments, respectively.
\end{enumerate}

\subsection{Experimental Results and Performance Analysis}
%The entire data transmission process is depicted in Fig. \ref{fig12}. To obtain the channel SNR, a random frame is transmitted to test the channel SNR and then feed it back to Jetson Xavier NX transmitter along with user requirements through the control channel. Afterwards, control information such as mask list and frame structure is transmitted to Jetson Xavier NX receiver through the control channel. Finally, as described in subsection B, the data frames are transmitted.

\begin{table*}[htbp]
  \renewcommand{\arraystretch}{1.5}
  \begin{center}
   \caption{Performance measured in real channels by changing antenna distance} 
    \label{tab2}
    \begin{tabular}{|c|c|c|c|c|c|} 
      \hline
      \textbf{Distance[m]} & \textbf{SNR} & \textbf{\makecell[c]{ASCViT-JSCC R\\PSNR+CS/SSIM+CS}} & \textbf{\makecell[c]{ASCViT-JSCC\\PSNR+CS/SSIM+CS}} & \textbf{\makecell[c]{DeepJSCC\\PSNR+CS/SSIM+CS}} & \textbf{\makecell[c]{BPG LDPC 16-QAM\\PSNR+CS/SSIM+CS}} \\
      \hline
      0.2 & High & 0.911/0.936 & 0.924/0.945 & 0.781/0.838 & \textbf{0.927/0.949}\\
      \hline
      1.4 & Medium & \textbf{0.837/0.845} & 0.803/0.852 & 0.639/0.764 & 0.443/0.458 \\
      \hline
      2 & Low & \textbf{0.431/0.457} & 0.286/0.358 & 0.221/0.274 & 0/0 \\
      \hline
    \end{tabular}
  \end{center}
  \vspace{-3mm}
\end{table*}
%\begin{table}[htb]
%  \renewcommand{\arraystretch}{1.5}
%  \begin{center}
%    \caption{Computational complexity of NNs.}
%    \label{tab3}
%    \begin{tabular}{|c|c|c|c|c|} 
      %\hline
      %\multicolumn{5}{c}{\textbf{Computational Complexity of NNs}} \\
%      \hline
%      \textbf{ASCViT-JSCC} & \textbf{\makecell[c]{Total\\FLOPs}} & 18.385G & \textbf{\makecell[c]{Total parameter\\number}} & 91.506M \\
%      \hline
%      \textbf{DeepJSCC} & \textbf{\makecell[c]{Total\\FLOPs}} & 0.628G & \textbf{\makecell[c]{Total parameter\\number}} & 0.140M \\
%      \hline
      %\vspace{1em}
%    \end{tabular}
%  \end{center}
%\end{table}
\begin{table}[htbp]
  \renewcommand{\arraystretch}{1.5}
  \begin{center}
    \caption{Inference time on the Jetson Xavier NX}
    \label{tab3}
    \begin{tabular}{|c|c|c|c|}
      %\hline
      %\multicolumn{4}{c}{\textbf{Inference Time of Practical Running on the Jetson Xavier NX}} \\
      \hline
      \textbf{\makecell[c]{ASCViT-JSCC\\encoding}} & 0.189s & \textbf{\makecell[c]{ASCViT-JSCC decoding/\\MAE decoding}} & 0.314s/0.112s \\
      \hline
      \textbf{\makecell[c]{DeepJSCC\\encoding}} & 0.005s & \textbf{\makecell[c]{DeepJSCC\\decoding}} & 0.004s \\
      \hline
      \textbf{\makecell[c]{BPG LDPC\\encoding}} & 3.210s & \textbf{\makecell[c]{BPG LDPC\\decoding}} & 0.924s \\
      \hline
    \end{tabular}
  \end{center}
  \vspace{-5mm}
\end{table}

The practical measurements from Scenario 1 are presented in Fig. \ref{fig13}. The results from real-world channels closely match those from simulated AWGN channels. ASCViT-JSCC trained at random SNRs outperforms its 10 dB counterpart, especially in low-SNR conditions, and consistently outperforms DeepJSCC across all SNRs. Minor differences between the simulated and real-world measurements may be attributed to hardware limitations, which are outside the scope of this study. These results demonstrate that the ICP is functioning as expected, and despite some differences in frame structure between the simulation and the ICP, the results remain comparable due to the simplicity of the testing scenarios.

Table \ref{tab2} presents the results from Scenario 2. ASCViT-JSCC trained at random SNRs outperforms other schemes in both medium and low-SNR environments. In high-SNR conditions, the traditional scheme still shows the best performance, although ASCViT-JSCC trained at 10 dB is only marginally behind. DeepJSCC consistently performs worse than ASCViT-JSCC at all SNR levels. The performance trends observed in Scenario 2 closely mirror those seen in both the simulation results and Scenario 1. In real-world wireless channels, the traditional scheme is prone to the \textit{cliff effect} when channel conditions deteriorate sharply, whereas NN-based schemes demonstrate graceful degradation in performance, with results comparable to or surpassing state-of-the-art separation-based digital schemes.

These experimental results confirm the advantages of ASCViT-JSCC in real-world conditions, establishing a solid foundation for further research in intelligent communications, including semantic communications. Future studies can leverage the ICP for practical evaluations, contributing to the development and standardization of intelligent communication systems.

\subsection{Complexity Analysis}
Table \ref{tab3} details the inference times for the networks running on the Jetson Xavier NX. No additional acceleration techniques were used, apart from the Jetson Xavier NX’s inherent GPU capabilities. ASCViT-JSCC takes approximately 0.189 seconds to encode an image, about 38 times longer than DeepJSCC. The decoding time for ASCViT-JSCC, including MAE decoding, is about 0.341 seconds, roughly 113 times longer than DeepJSCC. However, the traditional scheme (BPG LDPC) \cite{58,67} consumes significantly more time than NN-based schemes, as it relies on CPU-based processing.
 
Although the inference time for ASCViT-JSCC is currently longer than that of DeepJSCC, improvements in hardware and software optimization techniques are expected to reduce this gap. As high-performance computing devices and acceleration technologies continue to evolve, the trade-off between performance improvements and inference time in ASCViT-JSCC may become more favorable. In many applications, the performance gains offered by ASCViT-JSCC may justify the increased processing time.

%%%%%%%%%%%%%%%%  5 CONCLUSION %%%%%%%%%%%%%%%%%%%%%%
\section{Conclusion and Future Improvement}
In this paper, we proposed a novel scheme called ASCViT-JSCC for wireless image semantic transmission. Our approach differentiates between image components using YOLOv5 object detection, SIFT feature point detection and channel information, and leverages secondary parts---recovered by the MAE decoder at the receiver---to protect primary parts. Specifically, a ViT-based JSCC network with quantization modules was designed for end-to-end coding, ensuring efficient object preservation. In frequency-selective channels, CSIPA-Net was introduced to dynamically reallocate power based on CSI, further enhancing performance. Simulation results demonstrated the effectiveness of our scheme in preserving critical objects and improving reconstructed image quality.

To validate the practical feasibility of the proposed approach, we developed a OFDM prototype validation platform called ICP, which integrates embedded GPU systems and a SDR. By deploying various algorithms on the ICP, we obtained real-world measurements that verified the advantages of ASCViT-JSCC and allowed for a thorough complexity analysis. Our work, particularly the development of the ICP, provides a strong foundation for the future evolution and standardization of intelligent communications, including semantic communications.

Potential directions for enhancing the ICP include: (1) Developing MIMO support to expand research; (2) Aligning with 5G standards for more reliable data; (3) Enabling device mobility to test dynamic channels; and (4) Simplifying the interface to support various multimedia types simultaneously. 
    
\bibliographystyle{IEEEtran} 
\bibliography{ref} 

%\begin{IEEEbiographynophoto}{Jane Doe}
%Biography text here without a photo.
%\end{IEEEbiographynophoto}

%\begin{IEEEbiography}[{\includegraphics[width=1in,height=1.25in,clip,keepaspectratio]{fig1.png}}]{IEEE Publications Technology Team}
%In this paragraph you can place your educational, professional background and research and other interests.\end{IEEEbiography}

\end{document}